# Bitlis Eren Üniversitesi Fen Bilimleri Dergisi

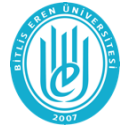

# Astrophysical Parameters of the Open Cluster Berkeley 6

Seliz KOÇ[1*], Talar YONTAN[2]

[1]*Istanbul University, Institute of Graduate Studies in Science, Programme of Astronomy and Space Sciences, 34116, Beyazıt, Istanbul, Turkey*

[2]*Istanbul University, Faculty of Science, Department of Astronomy and Space Sciences, 34119, Beyazıt, Istanbul, Turkey*
(ORCID: 0000-0001-7420-0994) (ORCID: 0000-0002-5657-6194)

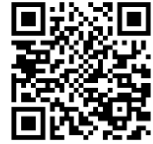

**Abstract**
In this study, the structural and basic astrophysical parameters of the poorly studied open cluster Berkeley 6 are calculated. Analyses of the cluster are carried out using the third photometric, spectroscopic, and astrometric data release of *Gaia* (*Gaia* DR3). The membership probabilities of stars located in the direction of the cluster region are calculated by considering their astrometric data. Thus, we identified 119 physical members for Berkeley 6. The colour excess, distance, and age of the cluster are determined simultaneously on the colour-magnitude diagram. We fitted solar metallicity PARSEC isochrones to the colour-magnitude diagram by considering the most probable member stars and obtained $E(G_{BP}-G_{RP})$ colour excess as 0.918±0.145 mag. The distance and age of the cluster are determined as $d$=2625±337 pc and $t$=350±50 Myr, respectively.

## 1. Introduction

Open star clusters (OCs) are groups of stars that are bound under weak gravitational forces and formed from the same molecular cloud that collapses under the same physical conditions. Therefore, the distance, metallicity, and age of the stars are similar, whereas their masses are different. So as the ages of OCs range from a few million to billions of years, they are essential objects to understand the star formation process and stellar evolution theories ([1]; [2]). OCs are located on the Galaxy disc. These characteristics make OCs important tools to understand the structure and dynamical evolution of the Galaxy disc (e.g., [3]; [4]; [5]).

The available information in the literature for the Berkeley 6 ($\alpha_{2000.0}$ = 01$^h$51$^m$12.7$^s$, $\delta_{2000.0}$ = 61°03'40"; $l$ = 130°.1008, $b$ = -00°.9760; [6]) is as follows: [7] analysed the CCD *UBVRI* observations of Be 6 and determined its structural and fundamental astrophysical parameters. They considered the radial density profile model of [8] and determined the core radius of the cluster as $r_c$ = 0.54±0.29 arcmin. By fitting theoretical isochrones of [9] to the observed (*B-V*) × (*V-I*) colour-colour diagram, [7] obtained the reddening of the Be 6 as $E(B-V)$ = 0.90±0.05 mag. Also, they shifted theoretical isochrones to the observed (*B-V*) × *V*, (*V-R*) × *V*, and (*V-I*) × *V* colour-magnitude diagrams and found the distance modulus, distance and age of Be 6 as $\mu_V$ = 14.80 mag, $d$ = 2.52±0.12 kpc and log $t$ = 7.5±0.1 yr, respectively. [10] improved an algorithm to obtain simultaneously reddening, distance, and age of clusters, and obtained parameters of Be 6 as $E(B-V)$=0.781 mag, $d$ = 2.5 kpc, and log $t$ = 8.6 yr, respectively. [6] estimated the distance and age of the Be 6 as $d$ = 3.05 kpc and log $t$ =8.35 yr using *Gaia* DR2 astrometric and photometric data. The mean proper-motion components of the cluster were calculated as ($\mu_\alpha \cos\delta$, $\mu_\delta$) = -0.857±0.076, -0.485±0.094 mas/yr by [6] (Table 1).

---

*Corresponding author: *seliskoc@gmail.com*

**Table 1:** Fundamental parameters for Berkeley 6 gathered from the literature. Columns denote the colour excess ($E(B-V)$), metallicity ([Fe/H]), distance moduli and distance ($\mu$, $d$), age ($t$), proper-motion components ($\mu_\alpha\cos\delta$, $\mu_\delta$), trigonometric parallaxes ($\varpi$) and reference. Errors of the parameters are shown in parenthesis.

| $E(B-V)$ (mag) | [Fe/H] (dex) | $\mu$ (mag) | $d$ (pc) | $t$ (Myr) | $\mu_\alpha\cos\delta$ (mas/yr) | $\mu_\delta$ (mas/yr) | $\varpi$ (mas) | Reference |
|---|---|---|---|---|---|---|---|---|
| 0.90 (0.05) | --- | 14.80 | 2520 (120) | 32 (8) | --- | --- | --- | [7] |
| 0.781 | --- | --- | 2500 | 400 | -1.29 | -3.22 | --- | [10], [11] |
| --- | --- | --- | 2950 (932) | --- | -0.857 (0.076) | -0.485 (0.094) | 0.31 (0.05) | [12], [13] |
| 0.581 | --- | 12.42 | 3051 | 224 | -0.857 (0.076) | -0.485 (0.094) | 0.31 (0.05) | [6] |
| 0.758 (0.032) | -0.182 (0.095) | --- | 2417 (212) | 110 (150) | -0.858 (0.090) | -0.495 (0.105) | 0.310 (0.048) | [14] |

## 2. Material and Method
### 2.1 Astrometric and Photometric Data

The third data release of *Gaia* (hereafter *Gaia* DR3, [17]) provides high quality astrometric and photometric data on more than 1.5 billion celestial objects. By using the equatorial coordinates of Be 6 given by [6] as the center of the cluster, we obtained 23,139 stars within a 20 arcmin radius with a magnitude range of $7 < G \leq 22$ mag. The obtained data from the catalogue are radius distance ($r$) from the centre, equatorial coordinates ($\alpha$, $\delta$), photometric magnitude and colours ($G$, $G_{BP}-G_{RP}$), The third data release of *Gaia* (hereafter *Gaia* DR3, [17]) provides high quality astrometric and photometric data on more than 1.5 billion celestial objects. By using the equatorial coordinates of Be 6 given by [6] as the center of the cluster, we obtained 23,139 stars within a 20 arcmin radius with a magnitude range of $7 < G \leq 22$ mag. The obtained data from the catalogue are radius distance ($r$) from the centre, equatorial coordinates ($\alpha$, $\delta$), photometric magnitude and colours ($G$, $G_{BP}-G_{RP}$), proper-motion components ($\mu_\alpha\cos\delta$, $\mu_\delta$) and trigonometric parallaxes ($\varpi$). The identification map of the open cluster Be 6 is shown in Figure 1.

### 2.2 Photometric Completeness Limits and Photometric Errors

To derive precise parameters, we identified the photometric completeness limit according to the $G$ magnitude of the stars. For this, we constructed star count versus $G$ magnitude histograms, as shown in Figure 2. In the figure, it can be seen that the star counts increase up to $G=21$ mag and start to decrease after this limit, which indicates that stellar incompleteness has set in. Thus, we adopted this value as the cluster photometric completeness limit and took into account the stars brighter than this limit for further analyses. The uncertainties of the *Gaia* DR3 photometric data were adopted as their interval errors. We calculated the mean errors of the $G$ magnitudes and $G_{BP}-G_{RP}$ colours of detected stars as a function of $G$ interval magnitudes and listed them in Table 2. Considering the adopted photometric completeness limit $G=21$ mag for Be 6, it can be seen from Table 2 that the mean internal errors for $G$ magnitudes and $G_{BP}-G_{RP}$ colours of the stars reach up to 0.007 and 0.146 mag, respectively.

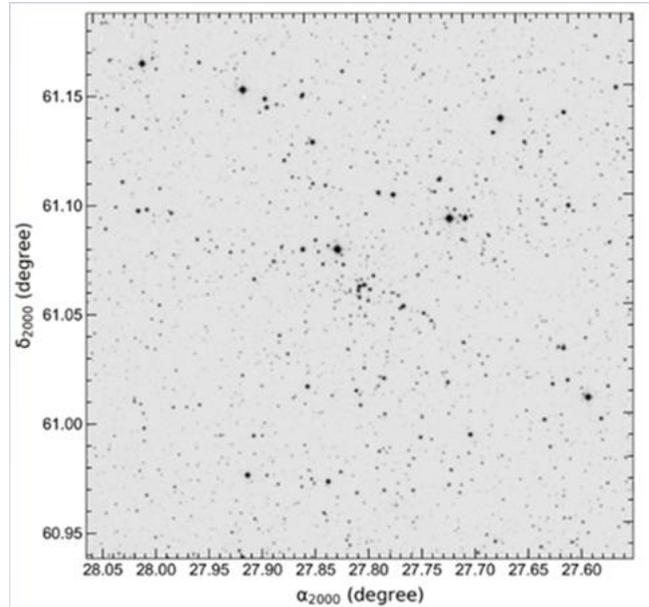

**Figure 1:** Identification chart of stars located in the region of Berkeley 6. Field of view of the optical chart is 20' × 20'. North and East correspond to the up and left directions, respectively.

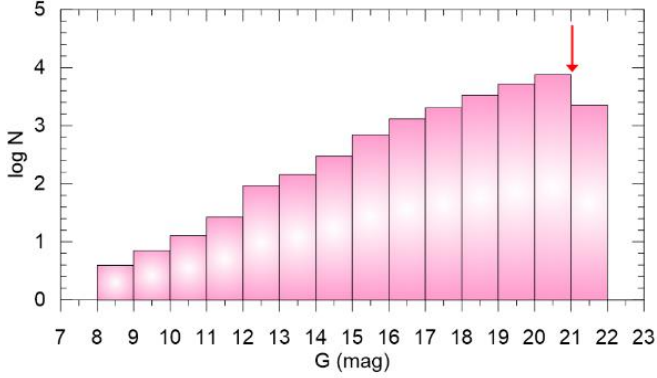

**Figure 2:** Histogram of number of stars according to *G* apparent magnitude intervals. The red arrow represents the faint limiting magnitude of the Berkeley 6.

**Table 2:** Mean photometric errors calculated for apparent *G* magnitude for Berkeley 6. *N* indicates the number of stars within the selected *G* magnitude intervals.

| *G* (mag) | *N* | $\sigma_G$ (mag) | $\sigma_{GBP-GRP}$ (mag) |
|---|---|---|---|
| ( 6, 12] | 51 | 0.003 | 0.005 |
| (12, 14] | 238 | 0.003 | 0.005 |
| (14, 15] | 301 | 0.003 | 0.005 |
| (15, 16] | 703 | 0.003 | 0.006 |
| (16, 17] | 1319 | 0.003 | 0.008 |
| (17, 18] | 2050 | 0.003 | 0.014 |
| (18, 19] | 3393 | 0.003 | 0.031 |
| (19, 20] | 5232 | 0.004 | 0.063 |
| (20, 21] | 7556 | 0.007 | 0.146 |
| (21, 22] | 2296 | 0.025 | 0.397 |

## 3. Results and Discussion

### 3.1 Structural Parameters of the Berkeley 6

To determine structural parameters such as the limiting and effective radii of the core region of Be 6, we carried out the Radial Density Profile (RDP) model given by [15]. For this, we first divided the cluster area into the series of concentric rings, which were adjusted to the central coordinates taken from [6]. Then we calculated the stellar densities (ρ) for each ring by dividing the number of stars by the ring area. We plotted stellar densities versus radius from the cluster centre (Figure 3) and fitted the RDP model of [15] to this distribution by following $\chi^2$ minimization. [15] described the model as $\rho(r) = f_{bg} + (f_0 / (1 + (r/r_c)^2))$, where *r* is the radius from the cluster centre, $f_{bg}$, $f_0$ and $r_c$ are the background stellar density, the central density and the core radius, respectively. Figure 3 shows the RDP of the cluster together with the best fitting [15] model for it. The best fit resulted in the core radius, background stellar density and central density, being $r_c$ = 0.501±0.029 arcmin, $f_{bg}$ = 10.705±0.269 stars/arcmin$^2$ and $f_0$ = 38.844±1.226 stars/arcmin$^2$, respectively. Also, we took into account the radius value at which stellar density is about to meet the background density (Figure 3 a grey horizontal line) as the cluster limiting radius ($r_{lim}$). This limit was adopted as $r$ = 5′ (3.82 pc) for Be 6, and we used only the stars inside this limiting radius in further analyses.

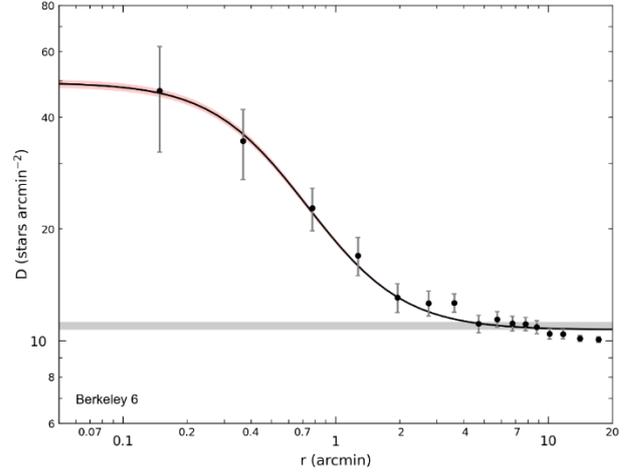

**Figure 3:** Radial density profile of Berkeley 6. Errors were derived using the expression of $1/\sqrt{N}$, where *N* represents the number of stars used in the density estimation. The solid line represents the optimal [15] profiles. The background density level and its errors are the horizontal grey bands. The King fit uncertainty (1σ) is shown with the red shaded region.

### 3.2 Membership Probabilities of Stars

Because the OCs are located near the Galactic plane, their members are highly affected by background star contamination. This makes it difficult to identify cluster members using only photometric data. Instead of this, the radial velocity and proper-motion values of stars are more useful in determining the membership. Astrometric and spectroscopic observations provide information on the radial velocity and proper-motion components of stars. However, spectroscopic observations of a great number of objects require much observing time, including larger telescopes. For this reason, the radial velocities of stars remain incapable of determining their membership. Because the components of OCs share a common origin, their movement vectors in the sky commonly point in the same direction. These properties make proper-motion components useful tools to separate cluster stars from field stars. Thanks to the astrometric data with high accuracy of *Gaia* DR3, the cluster members can be separated more precisely. In the study, we took into account *Gaia* DR3 proper-motion components as well as the trigonometric parallaxes of stars in the direction of Be 6 to calculate the membership probabilities of each. Membership analyses were utilized by using the UPMASK (Unsupervised Photometric Membership Assignment in Stellar Clusters; [16]) method. UPMASK is based on the clustering method, which is represented by *k*-means clustering, by detecting spatially concentrated groups and identifying the most likely cluster members. *k*-means is an integer number that varies from 5 to 25 and is not set directly by the user ([16]; [13]). [13] used *Gaia* DR2 astrometric data on 1,481 open clusters with the UPMASK methodology, and [6] successfully calculated the membership probabilities of stars in these clusters as a continuation of his work. We applied UPMASK by considering five-dimensional astrometric parameters of stars that contain positions (α, δ), proper-motion components ($\mu_\alpha \cos\delta$, $\mu_\delta$), trigonometric parallaxes ($\varpi$), and their uncertainties. During the application, we run 1000 iterations for Be 6. Membership probability is defined as the frequency with which a star is defined as a part of a clustered group. We reached the best result when the *k-means* value was set to 11 for Be 6. The possible cluster members were selected among the stars that were brighter than $G=21$ mag with membership probabilities $P \geq 0.5$ and located on the inside of the cluster limiting radius ($r = 5'$). These stars were adopted as physical members of Be 6. Thus, the number of member stars we reached for Be 6 is 119. These stars were used in the determination for the astrometric and astrophysical parameters of Be 6. In order to visualize the positions of the most probable member stars in Be 6, we constructed a vector-point diagram (VPD), as shown in Figure 4. It can be seen in the figure that Be 6 is embedded in the field stars. In Figure 4, the intersection of blue dashed lines represents the values of mean proper-motion components calculated from the most probable cluster members (119 stars with $P \geq 0.5$). These values were estimated as ($\mu_\alpha \cos\delta$, $\mu_\delta$) = (-0.894±0.004, -0.533±0.004) mas/yr, which are compatible with the results of all studies performed with *Gaia* observations for the Be 6 (see Table 1). To calculate the mean trigonometric parallax ($\varpi$) of Be 6 we plotted the parallax histogram considering the most probable cluster members and applied the Gaussian fit to the histogram of the selected stars (Figure 5). During the calculation of $\varpi$, we considered the stars with a relative parallax error ($\sigma_\varpi/\varpi$) of less than 0.2 to minimize uncertainties. We obtained the mean $\varpi$ of Be 6 as 0.32±0.03 mas. By applying the linear equation of $d$ (pc) = 1000/$\varpi$ (mas) to the mean trigonometric parallaxes, we reached the parallax distance as $d_\varpi$ = 3125±332 pc. Also, to compare results, we calculated the arithmetic mean ($\varpi_{mean}$) of trigonometric parallaxes from the most probable cluster members ($P \geq 0.5$) as $\varpi_{mean}$ = 0.325±0.034 mas, which is the same result as Gaussian fit. A histogram of the most probable stars, as well as the Gaussian fit (dashed black line) to the distribution are given in Figure 5.

### 3.3 Astrophysical Parameters of Berkeley 6

As the cluster members stars share similar physical properties and they contain a wide range of stellar masses, observational colour-magnitude diagrams (CMDs) are essential tools to visualize the morphology of open clusters, as well as determine their reddening, distances, and age.

To determine the metallicity, colour excesses, distance, and age of the Be 6, we fitted PARSEC isochrones of [18] to the observed CMD, considering the most probable cluster member stars. In the analyses, we fitted theoretical isochrones by eye, taking into account the most probable main-sequence and turn-off point members of the cluster. The CMD with the best fit isochrone is shown in Figure 6. We used the isochrones of log $t$ = 8.48, 8.54 and 8.60 yr. We considered solar metallicity value ([Fe/H] = 0 dex or $z$ = 0.0152) for relevant isochrones. The best fit isochrone provides colour excess of $E(G_{BP}-G_{RP})$ = 0.918±0.145 mag. Also, to compare this value with the literature, we used the equation of $E(G_{BP}-G_{RP}) = 1.41 \times E(B-V)$ given by [19] and calculated the *UBV* based colour excess as $E(B-V)$ = 0.651±0.103 mag. The errors for colour excesses were estimated by considering the mean photometric error of *Gaia* ($G_{BP}-G_{RP}$) colour in the $G$ completeness limit (see Table 2). The estimated $E(B-V)$ colour excess is compatible with the studies of [6], [10] and [14] within the errors. Moreover, the fitting procedure gives the age and distance moduli of Be 6 to be $t$ = 350±50 Myr and $(m-M_G)$ = 13.806±0.263 mag, respectively, which corresponds to the isochrone distance to be $d_{iso}$= 2625±337 pc (see Table 3). In addition to this, we calculated errors in distance moduli and distances using the relations presented in [22], [24], [25], [26].

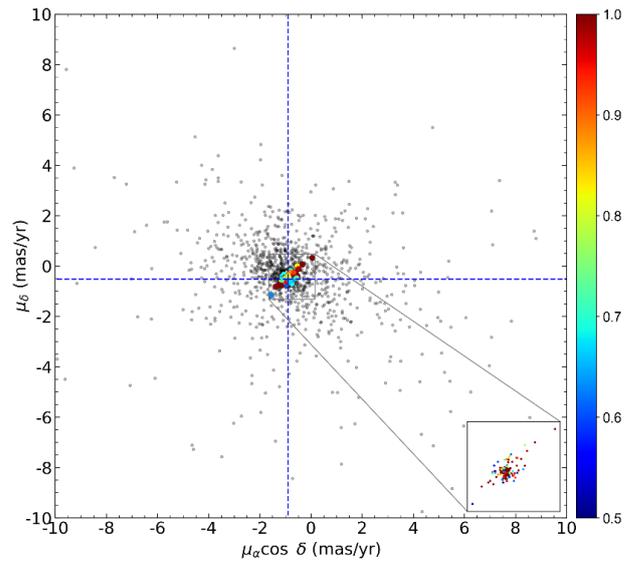

**Figure 4:** VPD of Berkeley 6 based on *Gaia* DR3 astrometry. The membership probabilities of the stars are showed with the colour scale shown on the right. In the zoomed frame represents the region of condensation for Berkeley 6 in the VPD. The intersection of the dashed blue lines is the point of mean proper motion of Berkeley 6.

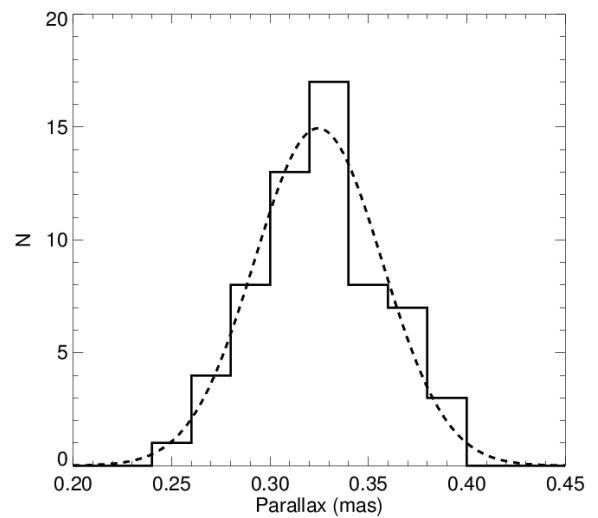

**Figure 5:** *Gaia* DR3 based trigonometric parallax histogram for Berkeley 6. The distribution of member stars with relative parallax error is less than 0.2 and the calculated Gaussian fit which shown with the dashed line.

These relations take into account the uncertainty in colour excess. The age obtained for Be 6 in this study ($t = 350\pm50$ Myr) is in good agreement with the results given by [10] and [11], whereas the isochrone distance value ($d_{iso} = 2625\pm337$ pc) is compatible with almost all the studies (except for the value of [6]) given by different researchers (see Table 1). The Galactocentric distance and Galactocentric coordinates were obtained as $R_{gc}$ = 9.90 kpc and $(X, Y, Z)_\odot$ = (−1691, 2008, -45) pc. These are compatible with the corresponding estimates in [12]. The basic parameters of Be 6 are listed in Table 3.

**Table 3:** Basic parameters of Berkeley 6 derived in this study.

| Parameter | Value |
|---|---|
| α (hh:mm:ss.s) | 01:51:12.7 |
| δ (dd:mm:ss.s) | 61:03:40.0 |
| $l$ (°) | 130.1008 |
| $b$ (°) | -0.9760 |
| $f_0$ (star/arcmin$^2$) | 38.844±1.226 |
| $r_c$ (arcmin) | 0.501±0.029 |
| $f_{bg}$ (star/arcmin$^2$) | 10.705±0.269 |
| $r_{lim}$ (arcmin) | 5 |
| $r$ (pc) | 3.82 |
| $\mu_\alpha \cos\delta$ (mas/yr) | -0.894±0.004 |
| $\mu_\delta$ (mas/yr) | -0.533±0.004 |
| Cluster members ($P \geq 0.5$) | 119 |
| $\varpi$ (mas) | 0.325±0.034 |
| $d_\varpi$ (pc) | 3077±322 |
| $E(G_{BP} - G_{RP})$ (mag) | 0.918±0.145 |
| $E(B-V)$ (mag) | 0.651±0.103 |
| [Fe/H] (dex) | 0 (assumed) |
| Age (Myr) | 350±50 |
| Distance module (mag) | 13.806±0.263 |
| Isochrone distance (pc) | 2625±337 |
| $X$ (pc) | -1691 |
| $Y$ (pc) | 2008 |
| $Z$ (pc) | -45 |
| $R_{gc}$ (kpc) | 9.90 |

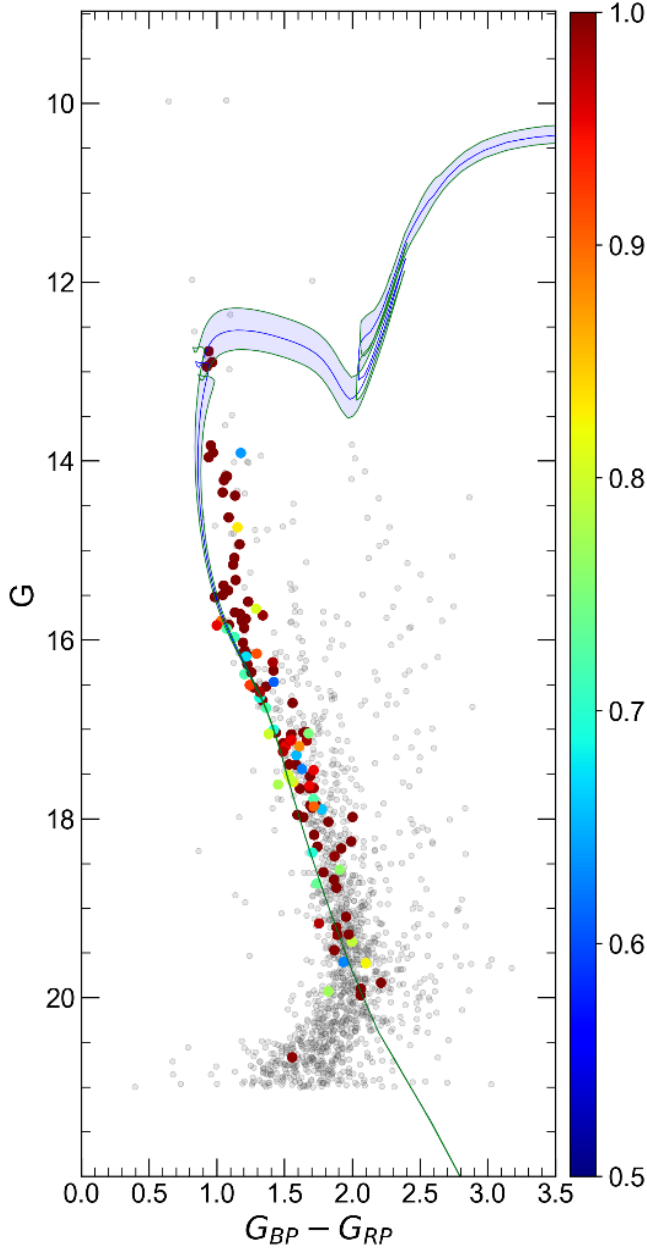

**Figure 6**: Colour-magnitude diagram of Berkeley 6. Different coloured circles show the membership probabilities according to the colour scales shown on the right side of the diagrams. Gray dots represent field stars. The blue lines show the PARSEC isochrones, while the shaded areas surrounding these lines are their associated errors.

## 4. Conclusion and Suggestions

In this study, we presented structural and astrophysical parameters of the Be 6 open cluster by using *Gaia* DR3 photometric and astrometric data. The astrometric parameters of *Gaia* DR3 were used to calculate the membership probabilities of stars located in the cluster region. The main results are given below:

- RDP analyses show that the limiting radius of the cluster is $r_{lim}$ = 5′ (3.82 pc).

- We classified 119 most likely cluster member stars with a membership probability $P \geq 0.5$ which are also lying within the limiting radius. The mean proper motion of Be 6 is estimated as $(\mu_\alpha\cos\delta, \mu_\delta)$ = (-0.894±0.004, -0.533±0.004) mas/yr.

- We considered the most likely member stars with a relative parallax error ($\sigma_\varpi/\varpi$) of less than 0.2 and determined the trigonometric parallax of the cluster as $\varpi$ = 0.325±0.034 mas. This value corresponds to the parallax distance of Be 6 to be $d_\varpi$ = 3077±322 pc.

- Astrophysical parameters were derived simultaneously by considering *Gaia* based CMD. PARSEC isochrones of solar metallicity [Fe/H] = 0 dex or $z = 0.0152$ were fitted through the most likely cluster members ($P \geq 0.5$). We took into account the morphology, such as the main-sequence, and turn-off point of the cluster, that is viewable on CMD during the fitting procedure. Hence, we determined the colour excess, age, and isochrone

- distance of Be 6 as $E(G_{BP}-G_{RP}) = 0.918\pm0.145$ mag, $t = 350\pm50$ Myr, and $d_{iso}= 2625\pm337$ pc, respectively.

- Although there is a difference of about 450 pc between the calculated parallax distance ($d_\varpi = 3077\pm322$ pc) and isochrone distance ($d_{iso}=2625\pm337$ pc) results, the two distance values are compatible within the errors. For objects far from the 2 kpc, the zero-point problem and observational biasness cause the measured *Gaia* trigonometric parallaxes to be smaller ([20];[21];[23]).

## Acknowledgment

We thank the anonymous referees for the insightful and constructive suggestions, which significantly improved the paper. This study is a part of the doctoral thesis of Seliz Koç. This study has been supported in part by the Scientific and Technological Research Council (TÜBİTAK) 122F109. We also made use of VizieR and Simbad databases at CDS, Strasbourg, France. We made use of data from the European Space Agency (ESA) mission Gaia[2], processed by the Gaia Data Processing and Analysis Consortium (DPAC)[3]. Funding for DPAC has been provided by national institutions, in particular the institutions participating in the Gaia Multilateral Agreement.

## Contributions of the Authors

Seliz Koç: Article writing, analysis.

Talar Yontan: Article writing, analysis.

## Conflict of Interest Statement

There is no conflict of interest between the authors.

## Statement of Research and Publication Ethics

The study is complied with research and publication ethics